\documentclass[preprint,12pt,times]{elsarticle}
\usepackage[breaklinks,colorlinks=true]{hyperref}
\usepackage{amsmath}
\usepackage{amssymb}
\usepackage{comment}

\newcommand{\Nc}{N_{\mathrm{c}}}
\newcommand{\Nf}{N_{\mathrm{f}}}
\newcommand{\calL}{\mathcal{L}}
\newcommand{\Tr}{\mathrm{Tr}}
\newcommand{\bmu}{\boldsymbol{\mu}}
\newcommand{\bA}{\boldsymbol{A}}
\newcommand{\bB}{\boldsymbol{B}}
\newcommand{\bj}{\boldsymbol{j}}
\newcommand{\br}{\boldsymbol{r}}

\journal{Physics Letters B}

\begin{document}
\begin{frontmatter}

\title{Magnetic enhancement of baryon confinement modeled via a deformed Skyrmion}
\author[1]{Shi Chen}
\author[1]{Kenji Fukushima}
\author[2]{Zebin Qiu}

\affiliation[1]{organization={Department of Physics, The University of Tokyo},
 addressline={7-3-1 Hongo, Bunkyo-ku},
 postcode={113-0033},
 city={Tokyo},
 country={Japan}}

\affiliation[2]{organization={Department of Physics, Keio University},
 addressline={4-1-1 Hiyoshi, Yokohama},
 postcode={223-8521},
 city={Kanagawa},
 country={Japan}}

\begin{abstract}
  We discuss the baryon properties under a strong magnetic field.  We adopt the Skyrme model and calculate the magnetic field dependence of the mass and the pressure distribution in the soliton.  We elucidate a magnetically induced contribution to the pressure sum rule and interpret it as an extra confining force.  We also quantize the soliton to estimate the difference between the proton and the neutron and find a simple relation between the pressure and the mass differences.
\end{abstract}


\end{frontmatter}

\section{Introduction}

Quantum chromodynamics (QCD) accommodates various phases with different physical degrees of freedom.  The elementary particles are quarks and gluons, and the physical states which can be experimentally observed are only color-singlet bound states of quarks and gluons; namely, the mesons and the baryons.
In the low-energy regime physical processes are dominated by dynamics of the lightest mesons, i.e., the pions realized as the (pseudo) Nambu-Goldstone bosons in the chiral broken phase.
The Chiral Effective Theory (ChEFT) is a systematic low-energy expansion of QCD in terms of the pions, particularly in the form of a non-linear sigma model with the target space $SU(\Nf)$, where $\Nf$ stands for the flavor number.
For modern nuclear physics, the ChEFT is one of the most successful \textit{ab initio} frameworks based on QCD; see, e.g., Ref.~\cite{Bogner:2009bt} for a review. 

Since the ChEFT postulates the spontaneous breaking of chiral symmetry and the confinement of quarks, there seems to be no way to probe the microscopic origin of these non-perturbative phenomena with the ChEFT calculations.  It is true for the mesonic sector; nevertheless, we may be able to diagnose the baryons.
The first hint comes from the large-$\Nc$ limit where $\Nc$ is the number of colors.
It is not difficult to understand that the large-$\Nc$ limit reduces QCD into a theory with planer diagrams only~\cite{tHooft:1973alw}, leading to a non-interacting theory in terms of mesons.  
The multi-body force between quarks is highly suppressed, which allows a Hartree-Fock analysis for the baryonic bound states~\cite{Witten:1979kh}.
Interestingly, such analysis found that baryons behaves the same as quantum solitons~\cite{Witten:1979kh}.
The second hint comes from 't Hooft anomaly matching.
The current algebra among the chiral flavor and the baryon currents calculated in massless QCD has an irremovable symmetry-violating contact term.
This is called 't Hooft anomaly and is argued to be preserved by the renormalization group flow.
If ChEFT is a sufficiently reliable infrared description, we should also see this anomaly in ChEFT\@.
Therefore, the ChEFT Lagrangian must include a Wess-Zumino-Witten (WZW) term~\cite{Witten:1983tw}, and the topological current for the homotopy group, $\pi_3(SU(\Nf))=\mathbb{Z}$, must be recognized as the baryon current~\cite{Witten:1983tx}.

Combining the two hints above strongly suggests that baryons arise in ChEFT in the form of topological solitons protected by $\pi_3(SU(\Nf))=\mathbb{Z}$.
It is well-known that the soliton energetically collapses in the lowest order of the ChEFT and Skyrme added a higher derivative term to circumvent the Derrick scaling theorem.
This formulation is called the Skyrme model~\cite{Skyrme:1961vq,Skyrme:1962vh},
which is founded as the simplest extension of the ChEFT;
in this sense, the Skyrme model is a model involving a parameter corresponding to the strength of the Skyrme term, and at the same time, it is a theory having a firm connection to QCD\@.
On top of the WZW term's capability of assigning fermionic statistics to the solitons after quantization,
the Skyrme model provides us with powerful non-perturbative approaches to nuclear properties.

Recently, in nuclear physics, the strong magnetic field is of increasing interest; see Ref.~\cite{Miransky:2015ava} for a comprehensive review.
Theoretically, the strongest magnetic field in the universe (up to the scale $eB\sim \mathrm{GeV}^2$) is expected in the high-energy collision of heavy ions in the laboratory, as numerically simulated in Refs.~\cite{Skokov:2009qp,Deng:2012pc,McLerran:2013hla}.
This strong magnetic field could be a probe to topological fluctuations in the QCD vacuum, which triggered intensive discussions of the chiral magnetic effect~\cite{Kharzeev:2007jp,Fukushima:2008xe} and similar anomalous transport~\cite{Kharzeev:2015znc}.
Such magnetically induced effects are also highly relevant to neutron star physics.
The surface magnetic field of the magnetar is of order $10^{11}$~Tesla, which is several orders smaller than the typical QCD scale, but it may still assist the realization of exotic phases such as the chiral soliton lattice system~\cite{Brauner:2016pko,Brauner:2019rjg}.
Motivated by these physics perspectives, the lattice-QCD simulations at the strong magnetic field have been studied with great care~\cite{DElia:2010abb,Bali:2011qj} and the results have revealed nontrivial responses of the QCD vacuum to the magnetic field, which includes the inverse magnetic catalysis at high temperature.
The hadron spectra have been also investigated with increasing magnetic field; see Ref.~\cite{Hidaka:2012mz} for scalar and vector mesons.
In particular, in Ref.~\cite{Ding:2020hxw}, the light meson properties even including neutral pions and kaons have been studied for $0.05<eB<3.35\,\mathrm{GeV}^2$, which reports that the masses of the neutral mesons are decreasing with increasing magnetic field.
This unexpected behavior of neutral mesons challenges our understanding of chiral symmetry breaking and confinement.
It is then a natural question to ask what would happen for baryons under a strong magnetic field.
In this paper, we answer this question by solving the Skyrmion deformed by the effect of the magnetic field.

Actually, in our preceding work~\cite{Chen:2021vou}, we have already solved this problem but our emphasis was put on a different aspect  of the dense nuclear matter which can be approximated by the Skyrme Crystal.
In this way, we have established a topological phase transition from normal nuclear matter to an exotic phase with the pion domain walls or the chiral soliton lattice.
We also addressed, but only slightly, the properties of an isolated baryon, i.e., the baryon exhibits prolate deformation with a major axis along the magnetic field.
In the present paper, we further pursue this direction, looking into the single baryon property.
We would like to stress that our physics arguments in this paper are quite general and based on the sum rule of the pressure that follows from the conservation law of the energy-momentum tensor.
Hence our choice of the Skyrme model is just for the purpose of demonstration whilst our physical results can be understood in a model-independent way.

We note that the energy-momentum tensor of a single baryon state is attracting interest in the context of the future electron-ion collider (EIC) experiment.
The form factors associated with the energy-momentum tensor can be decomposed, from which the D-term can be extracted; see Ref.~\cite{Polyakov:2018zvc} for a review, including the implication about the D-term~\cite{Polyakov:1999gs}.
For preceding works on the Skyrmion pressure, see also Ref.~\cite{Gibbons:2010cr} for a Skyrmion--anti-Skyrmion system.
Accordingly, as articulated in Ref.~\cite{Burkert:2018bqq}, the pressure distribution inside the proton, especially, the confining force from the outer shell which should be balanced with the repulsive pressure in the deep central region will be directly measured.
This is totally a novel direction to deepen our insight into the confinement problem.
In the present work, we will conclude that the baryons may be confined with the help of an extra confining force if they are put into an environment with a strong magnetic field.
This implies that the baryon confinement is assisted by an external magnetic field.
Our prediction of magnetic enhancement of baryon confinement can be in principle tested in the lattice-QCD simulation.
It is also worth noting that the energy-momentum tensor of the Skyrmion draws attention in the self-gravitating Einstein-Skyrme system, for which the energy-momentum tensor compatible with the metric can lead to topological solutions consistent with cosmological scenarios, and the solutions represent traversable wormholes~\cite{Canfora:2013osa,Ayon-Beato:2015eca}.
The physical contents are far different, but the technical treatments have similarities.
It would be interesting to compare the axisymmetric spacetime geometry in cosmology and our axially generalized hedgehog in the magnetic field.


\section{Skyrme model under a magnetic field}
\label{sec:model}

We briefly overview the Skyrme model in terms of the chiral field $\Sigma \in SU(\Nf)$ in the presence of the electromagnetic gauge field $A_\mu$.
The Lagrangian density $\calL=\calL_{0}+\calL_{\mathrm{WZW}}$ contains $\calL_{0}$, the ChEFT leading-order term with the Skyrme term, and $\calL_{\mathrm{WZW}}$, the gauged WZW term.
The former reads
\begin{align}
  \calL_{0} = -\frac{f_\pi^2}{4} \Tr(L_\mu L^\mu)
  + \frac{f_\pi^2 m_\pi^2}{2}\Tr(\Sigma-1)
  + \frac{1}{32a^{2}}\Tr([L_{\alpha},L_{\beta}][L^{\alpha},L^{\beta}])\,.
  \label{eq:Lag}
\end{align}
We introduced the $SU(\Nf)_{\mathrm{L/R}}$ current  as $L_\mu \equiv \Sigma^\dagger D_\mu\Sigma$
and $R_\mu \equiv \Sigma D_\mu\Sigma^\dagger$
with the covariant derivative $D_{\mu}\Sigma\equiv\partial_{\mu}\Sigma-iA_{\mu}\left[Q,\Sigma\right]$.
In our convention, $A_{\mu}$ absorbs the elementary charge $e$ and $Q$ is the dimensionless electric charge matrix which is $Q=1/6+\tau^{3}/2$ for $\Nf=2$.
We fix the model parameters as $f_\pi=54.0\,\text{MeV}$ and $a=4.84$ according to Ref.~\cite{Adkins:1983hy}
with the choice of the pion mass $m_\pi=138\,\text{MeV}$.

For $\Nf=2$, the gauged WZW term further comprises two parts, $\calL_{\mathrm{WZW}} = \calL_{\mathrm{WZW}}^0 + \calL_{\mathrm{B}}$.
The primitive WZW term $\calL_{\mathrm{WZW}}^0$ is the $\mathbb{Z}_2$ $\theta$-angle that assigns $-1$ to nontrivial configurations in $\pi_4(SU(\Nf))=\mathbb{Z}_2$ and $+1$ to trivial ones~\cite{Witten:1983tx}.
Such a term will determine the fermionic statistics of Skyrmions as we shall see in the quantization section.
The second part reads,
\begin{equation}
\calL_\mathrm{B}=qA_\mu j_{\text{B}}^\mu,
\end{equation}
which gauges the baryonic symmetry with a coefficient $q\equiv (\Nc/\Nf)\Tr Q=1/2$.
Here the baryon current is given by
\begin{equation}
 j_{\text{B}}^\mu = \frac{1}{24\pi^2} \epsilon^{\mu\nu\alpha\beta}
 \left\{ \Tr\left(L_\nu L_\alpha L_\beta\right)
 +i\frac{3}{2}F_{\alpha\beta}\Tr\left[Q
 \left(L_\nu - R_\nu\right)\right]\right\} \,.
\label{eq:jB}
\end{equation}
We note $\varepsilon^{0123}=\varepsilon^{0\rho\phi z}=\varepsilon^{0r\theta\phi}=-1$ in the Cartesian, the cylindrical, and the polar coordinates, respectively, in our convention.
The baryon number,
$N_{\text{B}}=\int d^3x\,j_{\text{B}}^0$, arises as
$\pi_3(SU(2))=\mathbb{Z}$,
which is intact for the Skyrmion even with
an external magnetic field, as antecedently discussed
in Ref.~\cite{Chen:2021vou}.

In this work, we focus on the energy-momentum tensor and derive it by the Noether theorem.
If we translate the dynamical field $\Sigma$ along an infinitesimal vector field $\epsilon^{\nu}$, i.e. $\delta\Sigma=\epsilon^{\nu}\partial_{\nu}\Sigma$, we shall find $\delta S=\int d^4x\,\partial_{\lambda}\epsilon_{\nu}T^{\lambda\nu} + \int d^4x\,\epsilon^{\nu}j_{\mathrm{Q}}^{\lambda}F_{\lambda\nu}$ with $j_{\mathrm{Q}}^{\,\mu} \equiv \delta\calL/\delta A_\mu$ being the electric current, and
\begin{align}
  T^{\mu\nu} \equiv -\frac{f_\pi^2}{2}\Tr(L^\mu L^\nu)
  + \frac{1}{8a^2}\Tr([L^\mu, L_\alpha] [L^\nu, L^\alpha])
  - g^{\mu\nu}\calL_{0}\,.
  \label{eq:Tcov}
\end{align}
Such $T^{\mu\nu}$ is the Noether current corresponds to the spacetime translation symmetry, i.e. the energy-momentum tensor.
We note that eventually $\calL_{WZW}$ does not enter the expression of Eq.~\eqref{eq:Tcov} but it gives a contribution to $j_{\mathrm{Q}}^{\,\mu}$.
The variation $\delta S$ should vanish on the solution to the equation of motion due to the stationary-action principle, which yields the conservation law assorted with a Lorentz force,
\begin{equation}
    \partial^{\lambda} T_{\lambda\nu} = j_{\mathrm{Q}}^{\,\lambda} F_{\lambda\nu}\,.
    \label{eq:Cons}
\end{equation}
If only to obtain the energy-momentum tensor $T^{\mu\nu}$ (without deriving the conservation law), one can just differentiate the action with respect to the metric, or translate the background gauge field $A_{\mu}$ together with $\Sigma$\footnote{
This last approach has a subtlety: We need to interpret what is ``translating'' a vector field along a vector field.
To get Eq.~\eqref{eq:Tcov}, we must adopt a geometric interpretation treating the translation as the infinitesimal diffeomorphism, i.e. the Lie derivative $\delta A_{\alpha}=\epsilon^{\nu}\partial_{\nu}A_{\alpha}+A_{\nu}\partial_{\alpha}\epsilon^{\nu}$.
\label{ft:T}
}. 

Our choice of gauge potential reads:
\begin{equation}
  A_0=0\,,\qquad
  \bA=\frac{B}{2}\boldsymbol{r}\times\hat{z}
  \qquad\left(B\geq0\right)\,,
\end{equation}
which means $\bB=-B\hat{z}$.
Here, our convention of $\bB$ is directed along the \textit{negative} $z$ axis.
Because the magnetic field breaks rotational symmetry,
we should generalize the original hedgehog Ansatz.
Specifically, we adopt the parametrization,
$\Sigma=\Pi_4 + i\boldsymbol{\tau}\cdot\boldsymbol{\Pi}$
with the Pauli matrices $\boldsymbol{\tau}$, where
\begin{equation}
  \begin{split}
  &\Pi_1 = \sin f\,\sin g\,\cos\varphi\,,
  \qquad\Pi_3 = \sin f\,\cos g\,,\\
  &\Pi_2 = \sin f\,\sin g\,\sin\varphi\,,
  \qquad\Pi_4 = \cos f\,.
  \end{split} 
\label{eq:parametrization}
\end{equation}
Here, $f$, $g$, and $\varphi$ are functions of $r$ and $\theta$
in the three-dimensional polar coordinates $(r,\theta,\phi)$.
These functions should satisfy the boundary conditions as~\cite{Chen:2021vou}
\begin{equation}
  f(r=\infty,\theta)=0\,,\quad
  f(0,\theta)=\pi\,,\quad
  g(r,\theta=0)=0\,,\quad
  g(r,\pi)=\pi\,.
  \label{eq:bc}
\end{equation}
In the limit of $B=0$, $f\to f(r)$ 
and $g\to \theta$ recover the hedgehog form.
It would be beneficial to review the symmetry of our model.
Given a sufficiently small quark mass, the continuous symmetry is approximate chiral symmetry, $SU(2)_{\mathrm{L}}\times SU(2)_{\mathrm{R}}$.
However, $\Sigma$ retains only the diagonal $SU(2)_{\mathrm{V}}$, which locks up the iso-rotation inextricably with the spatial rotation.
The magnetic field $\boldsymbol{B}$ further restricts such lock-up rotation to the $\tau_{3}$-component in the iso-rotation bases, as we will take a closer look when we discuss quantization later.
We shall turn to the discrete symmetries.  Then we see that the Lagrangian except $\calL_{\mathrm{WZW}}$ retains the symmetries of parity $\mathcal{P}: \Sigma(\br)\to\Sigma^\dag(-\br)$
(where $\dagger$ is needed due to the pseudo-scalar nature
of pions) and G-parity $\mathcal{G}=\mathcal{C}\,e^{i\pi I_2}$ composed
with charge conjugation $\mathcal{C}: \Sigma\to\Sigma^{\mathrm{T}}$
and iso-rotation $e^{i\pi I_2}: \Sigma\to\Sigma^{\ast}$ and
$A_{\mu}\to -A_{\mu}$.
The introduction of $\calL_{\mathrm{WZW}}$ violates G-parity due to 
the ``incompleteness'' of the $U(1)_{\mathrm{B}}$ gauging.

We can determine these functions, $f$, $g$, and $\varphi$, to minimize the energy functional,
$M=2\pi\int_0^\infty dr \int_0^\pi d\theta\, r^2 \sin\theta\, T^{00}$,
which we call the Skyrmion/soliton mass (i.e., the mass before quantization).
We can immediately show that $\varphi=\phi$, and
for $f$ and $g$, we employ the finite element method to
solve the Dirichlet problem.
For $B=0$ the problem reduces to a standard Skyrmion solved with the original hedgehog Ansatz.
Because the effective nucleon magnetic moment is negative (see later discussions for our sign convention),
the soliton mass $M(B)$ should be a decreasing function of $B$ as long as the linear approximation is justified.
With our generalized hedgehog Ansatz, we can continue solving $M(B)$ for larger $B$ beyond the linear regime.
In this way, we find that $M(B)$ features one global minimum
$M(B_0=3.60f_\pi^2 a^2)=61.6f_\pi a^{-1}=687\;\text{MeV}$ where
$\sqrt{B_0}=496\;\text{MeV}$ in the physical units.
For $B>B_0$, the soliton mass $M(B)$ grows with increasing $B$.
We will explain this behavior after quantization in later discussions.

\section{Confinement from the Pressure Sum Rule}
\label{sec:pressure}

We can derive that the behavior of $M(B)$ is dictated by
\begin{equation}
  \frac{\partial M}{\partial \bB} = - \bmu\,,
  \label{eq:muB}
\end{equation}
where $\bmu$ can be readily recognized as the magnetic moment since 
\begin{equation}
  \bmu \equiv \int d^3 x\, \frac{1}{2} \br\times\bj_{\mathrm{Q}}\,.
  \label{eq:mu}
\end{equation}
We do not take the $B=0$ limit in Eq.~\eqref{eq:muB}.
Also, $\bmu$ is not yet the nucleon magnetic moment before the soliton is quantized.
Multiplying $x_\mu$ on both sides of the conservation law Eq.~\eqref{eq:Cons} and performing spatial integration by part under the condition that $x_\mu T_{\lambda\nu}$ vanishes sufficiently fast at the infinity, we arrive at the following sum rule:
\begin{equation}
  \int d^3 x\, T_{\mu\nu}
  = -\int d^3 x\, x_\mu j_{Q}^\lambda F_{\lambda\nu}\,.
\end{equation}
From this sum rule, it is straightforward to confirm that the spatial integration of the longitudinal pressure, $P_z$, is zero,
\begin{equation}
  P_z = \int d^3 x\, p_z = \int d^3 x\, T_{zz} = 0\,.
  \label{eq:sum_pz}
\end{equation}
This is a widely known relation~\cite{Polyakov:2018zvc} and
we can immediately give a clear physical interpretation.
In the vicinity of the soliton center, $p_z$ is supposed to be outward-directed (which we define as \textit{positive} direction) to prevent the system from collapsing.
Meanwhile, for the system not to explode, $p_z$ must turn inward-directed, i.e., \textit{negative} in our sign convention, near the surface of the soliton.
Then, the outward force from quark motions and the inward force from quark confinement should be balanced for the self-bound system of soliton, and Eq.~\eqref{eq:sum_pz} is nothing but the balance equation of confinement.
In view of the Landau quantization exerted only on the transverse directions, it is natural that the balance equation for $p_z$ parallel to $\bB$ is unaffected.
By contrast, the balance of confinement on the transverse plane is significantly altered by the magnetic field as
\begin{equation}
  P = \int d^3 x\, p = \int d^3 x\, \frac{1}{3}\sum_{i=x,y,z}T_{ii}
  = -\frac{2}{3}\bmu\cdot\bB\,.
  \label{eq:sum_p}
\end{equation}
Here, we defined the rotationally averaged pressure $p$.
As mentioned before, $M(B)$ first decreases and then gets to increase as a function of $B$, which corresponds to positive (and negative) $\bmu\cdot\bB$ for small (and large, respectively) $B$.
We can understand that $\bmu\parallel \bB$ reduces the magnetic energy,
$E_{m}=-\bmu\cdot\bB$, so that $P<0$ can be expected for small $B$.
The interpretation for $P<0$ is that the outer shell region must have a stronger confining force.
In other words, $\bmu\cdot\bB>0$ tends to disfavor confinement and extra inward pressure is required.
In contrast, for $B>B_0$, the Skyrmion is squeezed into a smaller transverse radius due to heavy $\pi^\pm$, i.e., prolate deformed, as we demonstrated in our previous work~\cite{Chen:2021vou}.
This transverse shrinking results in a stronger repulsive force with $P>0$.
Because less confining pressure can saturate the balance condition, we can interpret Eq.~\eqref{eq:sum_p} with $\bmu\cdot\bB < 0$ as enhanced confinement assisted by the magnetic effect.

\begin{figure}
  \minipage{0.5\textwidth}   
  \includegraphics[width=\linewidth]{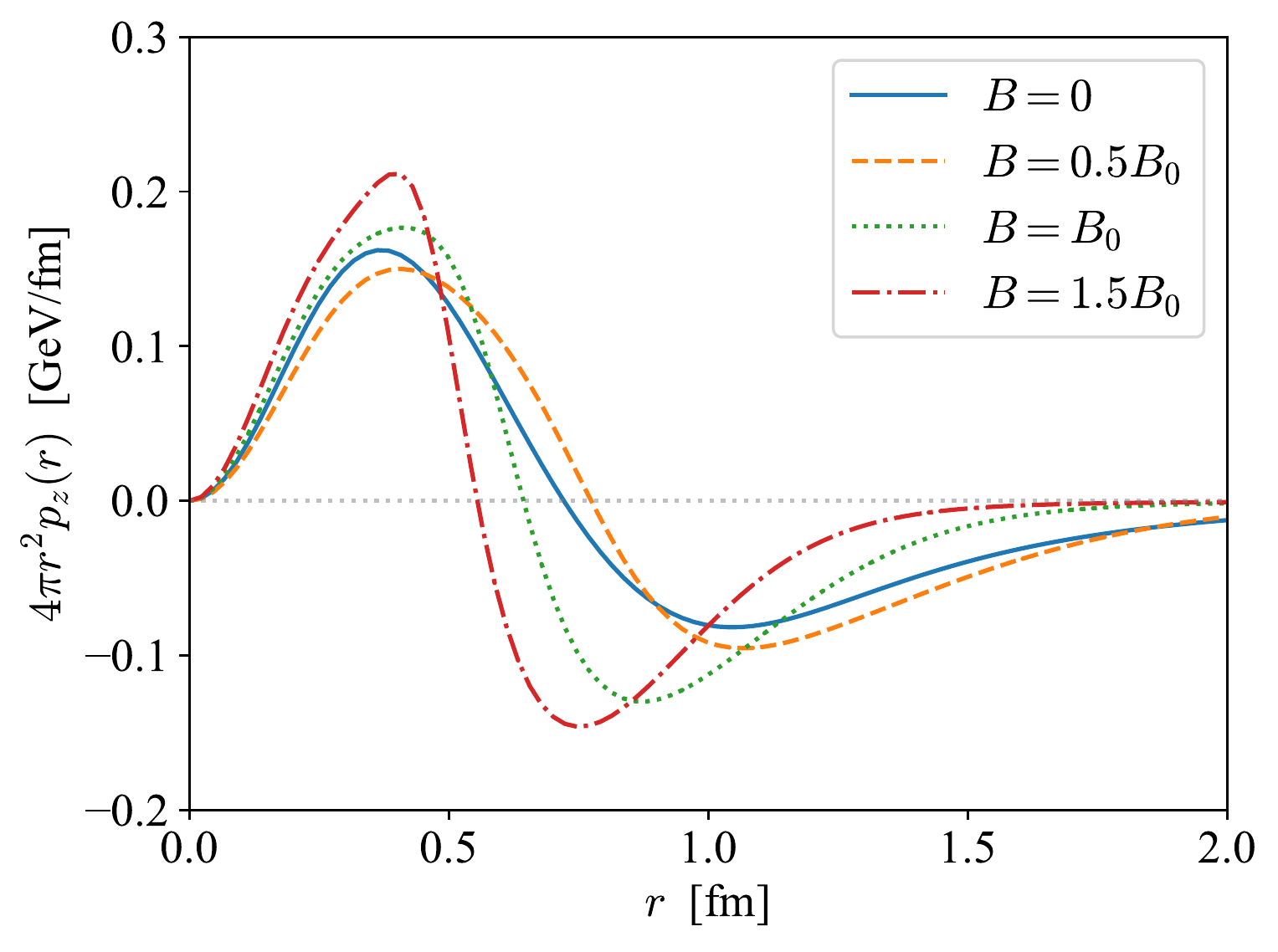}   
  \endminipage\hfill 
  \minipage{0.5\textwidth}
  \includegraphics[width=\linewidth]{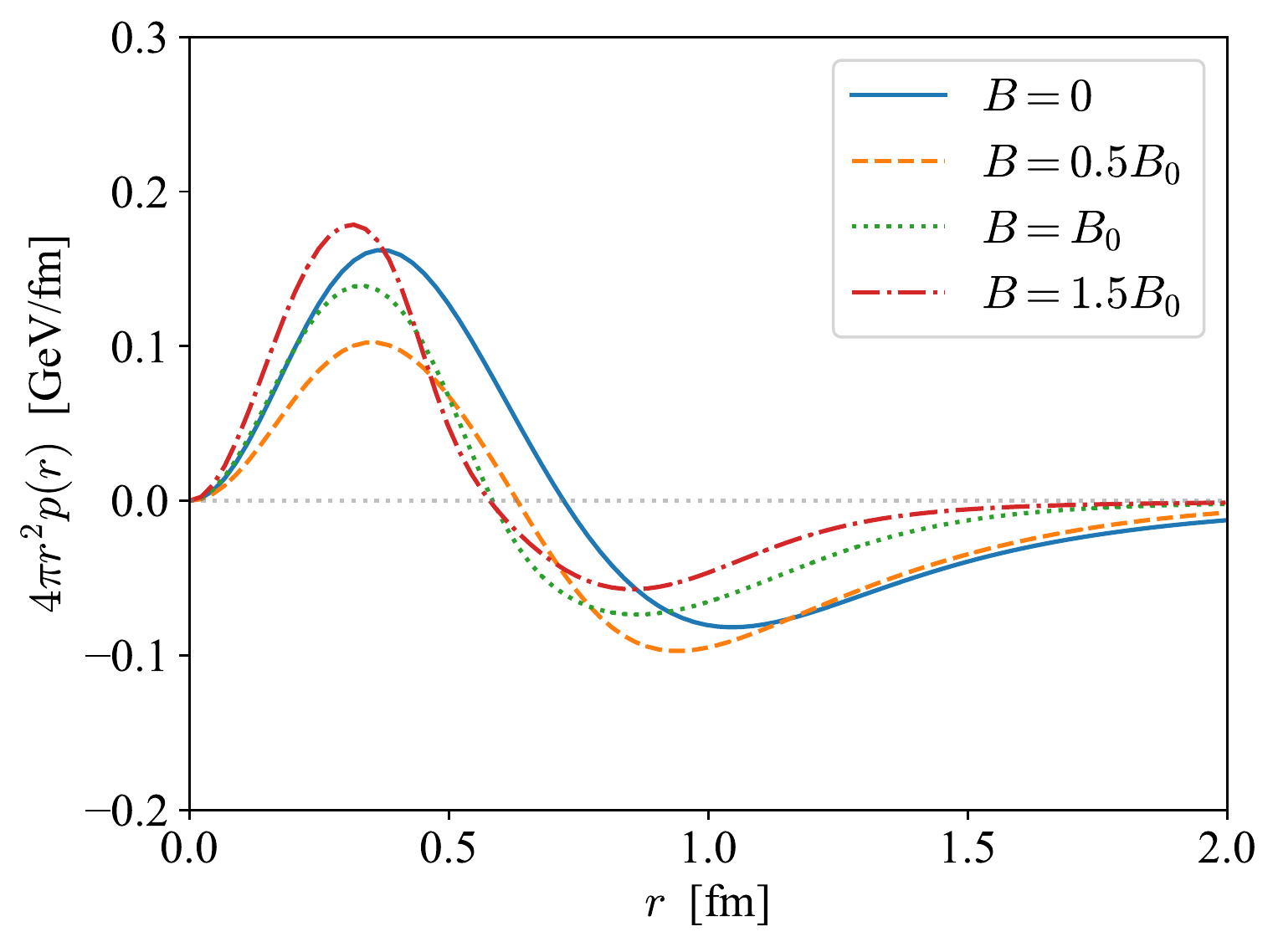}   
  \endminipage\hfill 
  \caption{Longitudinal pressure $p_z$ (left) and rotationally averaged pressure $p$ (right) after the angular integration multiplied by $4\pi r^2$.  For $B=B_0$, $|\bmu|=0$ is realized and the positive area with $p>0$ and the negative area with $p<$ are exactly balanced in both $4\pi r^2 p_z(r)$ and $4\pi r^2 p(r)$.}
  \label{fig:pressure}
\end{figure}

Let us take a closer look at the local pressure distribution inside the soliton.
The left panel in Fig.~\ref{fig:pressure} shows the angular-averaged longitudinal pressure, i.e.,
$p_z(r)=\frac{1}{2}\int_0^{\pi}d\theta\sin\theta\,T_{zz}(r,\theta)$, multiplied by $4\pi r^2$, so that the positive area and the negative area should be the same and the $r$ integration is vanishing according to the sum rule~\eqref{eq:sum_pz}.
We can indeed confirm that the sum rule holds for any magnetic field in the figure, and the stronger magnetic field tends to squeeze the pressure distribution to smaller spatial region.  
For the rotationally averaged pressure, the sum rule should be Eq.~\eqref{eq:sum_p}, and this is the case in the right panel in Fig.~\ref{fig:pressure}.
In the weak magnetic region, $0<B<B_0$ (where $B_0$ is defined for $|\bmu(B_0)|=0$ as discussed earlier), the right-hand side in Eq.~\eqref{eq:sum_p} is negative, and the dashed curve for $B=0.5B_0$ in the figure certainly has a larger negative area.
For $B=B_0$, the magnetic moment is vanishing, and as seen by the dotted curve in the figure, the total area becomes vanishing, while the positive area is enhanced in the dot-dashed curve for $B=1.5B_0$ in the figure.
For $B>B_0$ the peak positions are close for $4\pi r^2 p_z(r)$ and
$4\pi r^2 r(z)$, and the repulsive cores are squeezed by the strong magnetic field in a similar way.
However, the tail behaviors associated with confining shells are different, and this is because quark confinement is assisted by the Landau quantization which generates the energy gap $\sim \sqrt{B}$ among transverse orbits.

Here, we shall comment on a subtlety in the energy-momentum tensor.
Our $T_{\mu\nu}$ given in Eq.~\eqref{eq:Tcov} is obviously symmetric with respect to $\mu$ and $\nu$ and also gauge invariant, but we could have write down another form of the ``canonical'' energy-momentum tensor,
\begin{equation}
  \tilde{T}_{\mu\nu}
  =\frac{\delta\calL}{\delta(\partial^\mu \Sigma)}\partial_\nu\Sigma-g_{\mu\nu}\calL_0\,.
\end{equation}
This familiar form is somehow less ``canonical'' in our situation because to obtain $\tilde{T}^{\mu\nu}$ we need to add extra terms to $\delta\Sigma=\epsilon^{\nu}\partial_{\nu}\Sigma$ during applying the Noether theorem.
Thus in our theory with background gauge fields, our $T^{\mu\nu}$ in Eq.~\eqref{eq:Tcov} is the rather ``canonical'' one.
Also consequently, $\tilde{T}_{\mu\nu}$ does not respect the conservation law in the form of Eq.~\eqref{eq:Cons}.
To see this, we can compute the difference,
$\tilde{T}_{\mu\nu}-T_{\mu\nu}$, from which
we can quantify the (averaged) pressure difference,
\begin{align}
  \tilde{p}-p = \frac{1}{3}\Delta T^{\phi\phi}
  = \frac{1}{6}B\gamma \left( 1 - \frac{1}{2}Br^2\sin^2\theta\right)\cos f\,,
 \label{eq:deltap}
\end{align}
where we defined,
\begin{equation}
  \gamma = \sin^2f \sin^2g
  \left[f_\pi^2 +\frac{1}{a^2} \left(|\boldsymbol{\nabla}f|^2 + |\boldsymbol{\nabla}g|^2 \sin^2 f \right)\right]\,,
  \label{eq:gamma}
\end{equation}
which is the integrand for the moment of inertia we will soon encounter.
We note that the pressure difference~\eqref{eq:deltap} is nonvanishing even after the volume integration,
which means that the conservation law~\eqref{eq:Cons} is indeed modified.
Thus, for our discussions about confinement based on the pressure balance, our choice of the symmetric $T^{\mu\nu}$ is more preferable.

\section{Distinguishing the Proton and the Neutron}
\label{sec:quantization}

To identify our soliton with the proton/neutron,
we need to perform the projection of the soliton to the eigenstate of spin and isospin.
To this end, we vary the already attained static solution
$\Sigma$ as
\begin{equation}
  \Sigma(t) = e^{i\alpha(t)Q}\, \Sigma\, e^{-i\alpha(t)Q}
\end{equation}
with the collective coordinate $\alpha(t)$ that encodes the
time dependence.
Inserting this $\Sigma(t)$ into $\calL$,
we find the effective Lagrangian for $\alpha(t)$, i.e.,
\begin{equation}
  \calL = -M - \Phi\dot{\alpha}
  +\frac{1}{2}\Gamma\, \dot{\alpha}^{2}\,.
  \label{eq:Lalpha}
\end{equation}
Here, $M$ is the soliton mass, and the term $\propto \dot{\alpha}^2$ represents the rotational energy with the moment of inertia $\Gamma=\int d^3x\,\gamma$ where $\gamma$ is given in the above Eq.~\eqref{eq:gamma}.
The linear term in $\dot{\alpha}$ is rooted
in the WZW action.
The coefficient $\Phi$ has the physical
meaning of a transverse magnetic flux based on its explicit expression,
\begin{equation}
  \Phi = \frac{qB}{2}\int d^3 x\, (r\sin\theta)^2 \rho_T\,,
  \label{eq:Phi}
\end{equation}
where $\rho_T$ is the topological part of the baryon density independent of the gauge field, that is,
\begin{align}
  \rho_T  = \frac{\varepsilon_{ijk}}{24\pi^2} \mathrm{Tr}(L_i L_j L_k )
  = \frac{\sin^2 f \sin g}{2\pi^2 r^2 \sin\theta}
  (\partial_\theta f\,\partial_z g
  -\partial_z f\,\partial_r g)\,.
\end{align}
In Eq.~\eqref{eq:Phi} we see that a transverse area occupied by the soliton is effectively quantified by the weight $\rho_T$,
and the multiplication of $B$ makes an effective magnetic flux.
It should be noted, however, that
$\Phi$ is not simply proportional to $B$, but it shows a nonlinear saturation for large $B$.
Also, $\Gamma$ turns out a nonmonotonic function of $B$.

\begin{figure}  
  \minipage{0.5\textwidth}
  \includegraphics[width=\linewidth]{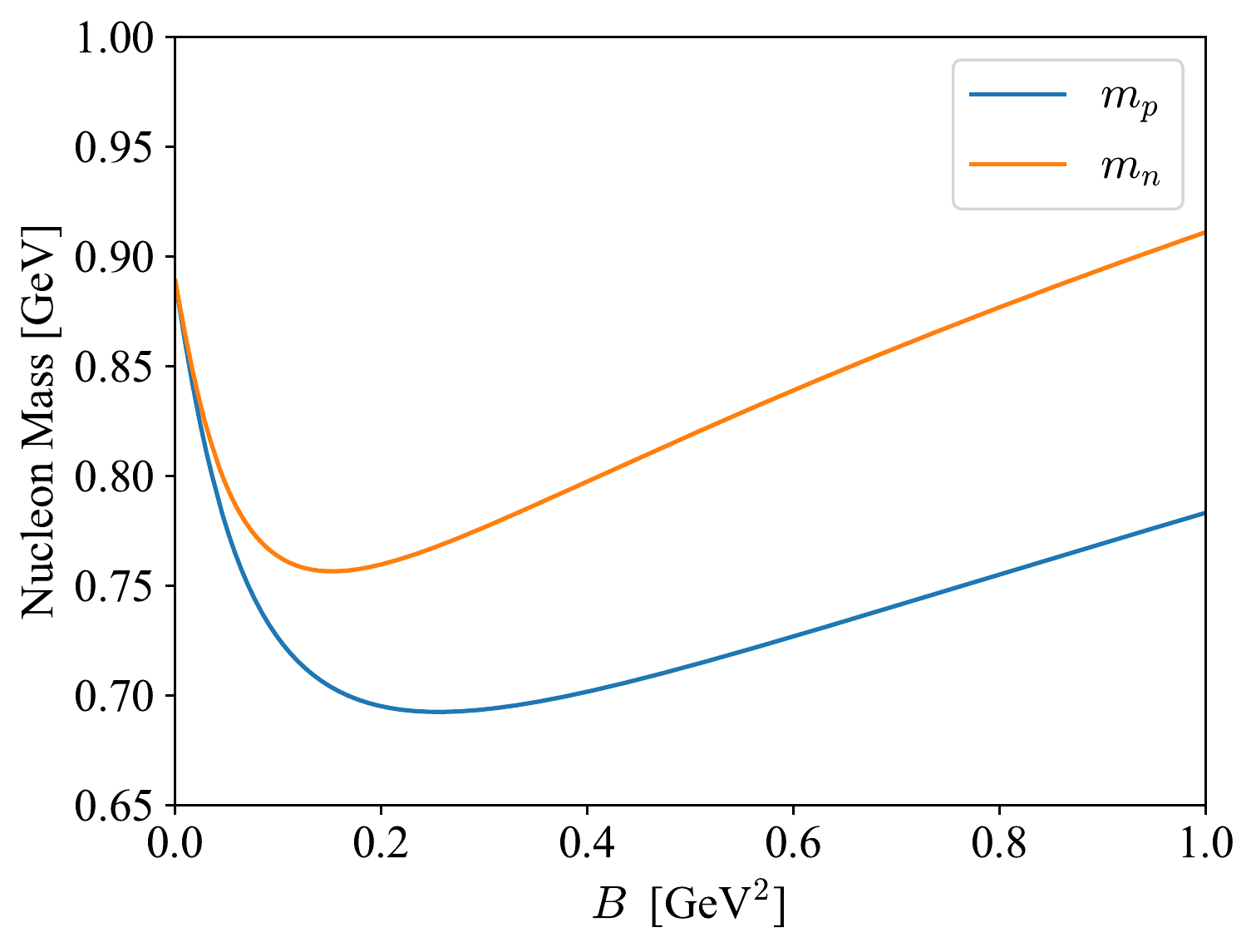}
  \endminipage\hfill
  \minipage{0.5\textwidth}
  \includegraphics[width=\linewidth]{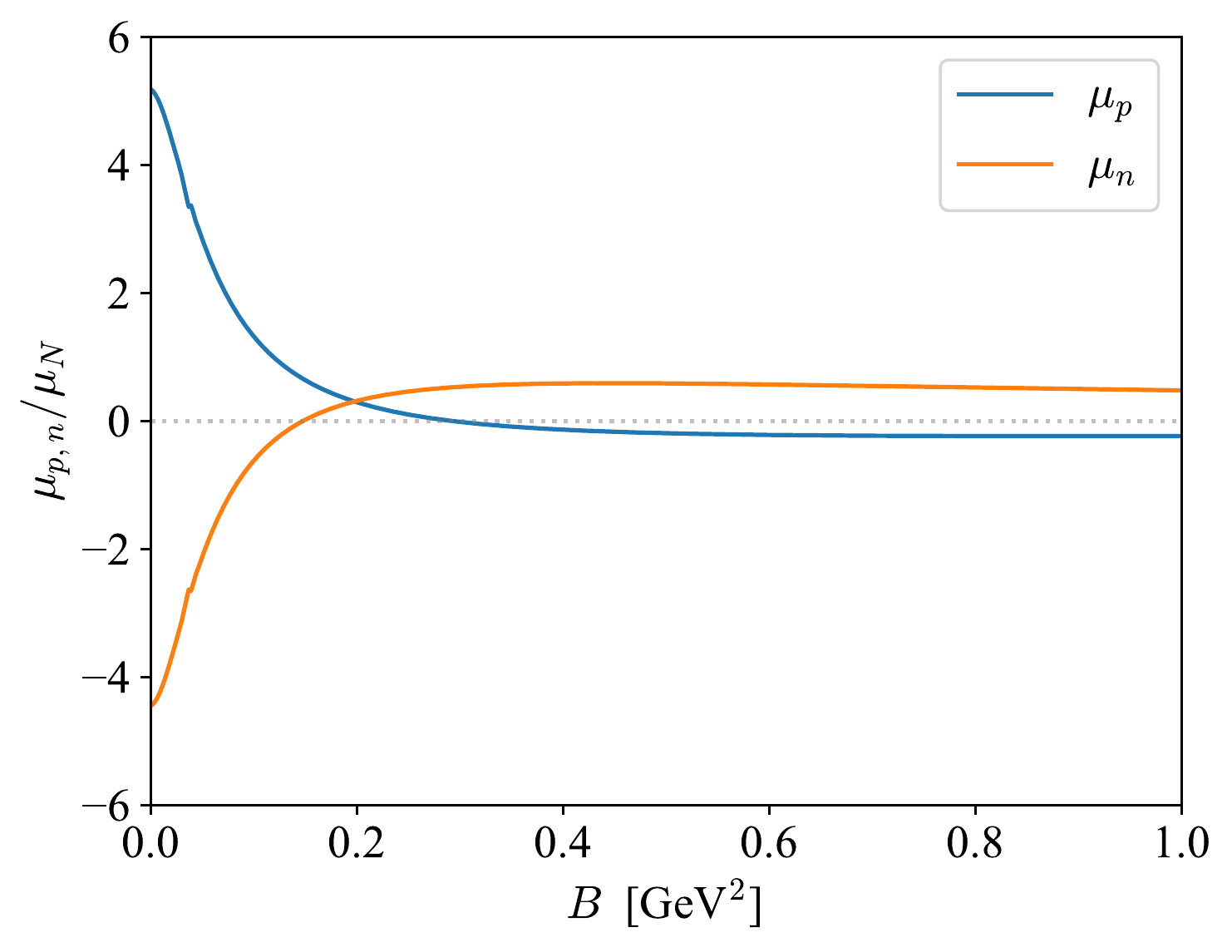}
  \endminipage\hfill
  \caption{Mass difference after quantization, $m_n-m_p$, under the strong magnetic field (left).  The magnetic moments of the proton and the neutron, $\mu_{p,n}$ in the unit of the nuclear magneton, $\mu_N=e\hbar/(2m_N)$ (right).}
  \label{fig:Bdiff}
\end{figure}

The Hamiltonian thus reads,
\begin{equation}
  H = \beta\dot{\alpha}-\calL
  = M + \frac{1}{2\Gamma} (\beta+\Phi)^2\,,
  \label{eq:H}
\end{equation}
where $\beta$ is the canonical momentum:
$\beta\equiv \delta\calL/\delta\dot{\alpha} = \Gamma\dot{\alpha}-\Phi$.
The primitive WZW term, i.e. the $\mathbb{Z}_2$ $\theta$-angle, dresses the amplitude of a soliton rotating by $2\pi$ with an extra
factor $(-1)^{\Nc}$.
Thus, $\Nc=3$ dictates the fermionic statistics, leading to a half-integer spectrum, i.e.,
\begin{equation}
  \beta = \frac{2n-1}{2}\,,\qquad n\in\mathbb{Z}\,.
\end{equation}
In view of the Noether theorem with regard to the isospin rotation in the $\tau_{3}$ component and the locked spatial rotation in the $\hat{z}$ direction,
we infer the quantum numbers accordingly,
$I_3 = -J_3 = -\beta$.
Provided that $\Phi,\Gamma\geq0$ and $\Phi$ is small for small $B$, 
we can find that the ground state has $\beta=-1/2$ to minimize $H$.  This means that
the ground state is $|p\!\downarrow\rangle$
and the first excited state is
$|n\!\uparrow\rangle$.  Then, from the expression of $H$,
we can immediately deduce the mass difference
between $|p\!\downarrow\rangle$ and $|n\!\uparrow\rangle$.
Specifically,
from the masses defined by
$m_n=H(\beta=1/2)$
and $m_p=H(\beta=-1/2)$,
we see the neutron gets heavier due to the magnetic effect by
$\Delta m=m_n-m_p=\Phi/\Gamma$ as reported in our previous work~\cite{Chen:2021vou}.  The left panel in Fig.~\ref{fig:Bdiff} shows the mass splitting between the proton and the neutron with increasing $B$.  As we mentioned, the overall structure which hits a minimum around $B\sim 0.2\,\text{GeV}^2$ inherits from the soliton mass behavior.

Here, an important remark is that we limit ourselves to the full polarization case for simplicity.  In the absence of the magnetic field, the rotational energy spectra should involve not $I_3$ and $J_3$ but the angular momenta squared like $I(I+1)$ and $J(J+1)$.
If the magnetic field is strong enough, it is expected that the rotation is eventually restricted to the axisymmetric motion.
Therefore, our treatment of quantization implicitly presumes strong $B$, while the soliton itself reduces to the hedgehog form in the $B\to 0$ limit.

Now, we shall discuss the pressure difference between
$|p\!\downarrow\rangle$ and $|n\!\uparrow\rangle$ as quantized Skyrmions in the magnetic field.
The nucleon magnetic moment $\mu_{n,p}$ is derived from the definition in Eq.~\eqref{eq:mu},
which can be conventionally decomposed into the isoscalar and isovector
parts associated with the baryon current $\bj_{\mathrm{B}}$
and the isospin current $\bj_{\mathrm{I}}$, respectively.
The magnetic moment after quantization is
\begin{align}
  \bmu_{p,n} & =\int d^{3}x\,\frac{1}{2} \boldsymbol{r}\times (q\bj_{\mathrm{B}} + \bj_{\mathrm{I}})\notag \\
  &= -\left[\frac{1}{2}\int d^3 x\,\gamma
  \left(1-\frac{1}{2}B r^2\sin^2\theta\right)
  -\frac{\left(\beta+\Phi\right)\Phi}{\Gamma B}\right]\hat{z}\,,
\end{align}
with $\beta=1/2$ for $|p\!\downarrow\rangle$ and $\beta=-1/2$ for $|n\!\uparrow\rangle$ .
The right panel in Fig.~\ref{fig:Bdiff} is the plot for $\mu_{p,n}$ which are the projected expectation value along the spin direction of either proton or neutron.  Because $|p\!\downarrow\rangle$ and $|n\!\uparrow\rangle$ have the opposite spin alignment, the magnetic moments should flip the sign accordingly, i.e.,
$\mu_p>0$ and $\mu_n<0$ in accord with experiments.
The values of $\mu_{p,n}$ in the figure look nearly three times larger than the empirical values in Ref.~\cite{Adkins:1983hy}, and this discrepancy is explainable from our approximation of full polarization (see also, e.g., Ref.~\cite{Ohtani:2004aw} in which a similar approximation was adopted).
Conclusively, by such an effective nucleon magnetic moment $\bmu_{p,n}$ we yield the nucleon pressure from the sum rule,
$P_{p,n} = -(2/3)\bmu_{p,n}\cdot\bB$.
As we have demonstrated earlier, $\bmu$ becomes
anti-parallel to $\bB$ (i.e., positive along $\hat{z}$)
for large magnetic fields.
The magnetic energy is thereby
increasing with $B$, supporting the confining feature against $P_{p,n}>0$.
This trend is unchanged after quantization, as we can easily make sure from the behavior of $m_p$ and $m_n$ in Fig.~\ref{fig:Bdiff}.
In this way, we can quantify the discrepancy in spatially integrated pressure between $|p\!\downarrow\rangle$ and
$|n\!\uparrow\rangle $ as
\begin{equation}
  P_n - P_p = \frac{2}{3} \frac{\Phi}{\Gamma}
  = \frac{2}{3} \Delta m\,.
\end{equation}
This relation appears consistent with our intuition.
Since the neutron enjoys a stronger $B$-induced assistance for confinement, the neutron can store more energy inside and thus become heavier.


\section{Discussions and Conclusions}
\label{sec:conclusion}

This work extends the analysis in our previous work where the technical formulation was established~\cite{Chen:2021vou}.
The major update is that we have taken into account the finite pion mass effect.
The overall qualitative behavior is hardly changed by $m_\pi\neq 0$, but it may be useful to clarify some differences.
In our previous work~\cite{Chen:2021vou},
we observed that the $B$-induced deformation on the soliton starts from the oblate direction for $B\lesssim 0.4f_\pi^2 a^2$.
Then, for $B\gtrsim 0.4f_\pi^2 a^2$, the deformation changes into the prolate direction.
In contrast, in the present study with $m_\pi=138\,\text{MeV}$, there is no such turnover in the deformation and the prolate shape is always favored.
The profile is optimized to minimize the energy or the integration of $T^{00}$.
The pion mass term enters $T^{00}$ via $f_\pi^2 m_\pi^2 (1-\cos f)$,
which repels $\pi^{0} \propto \sin f$ to an outer torus region inside the soliton.
This $\pi^0$ surrounding from the outer torus would prevent $\pi^\pm$ from expanding transversely for a weak $B$, which makes a contrast to the argument in Ref.~\cite{Chen:2021vou}.

An intriguing feature is also recognized in the soliton mass $M$ as a function of $m_\pi$.
We have numerically verified that $M$ grows up with increasing $m_\pi$ monotonically.
The growth rate, however, becomes slower for larger $B$.
This is easily understood through the structure in $T^{00}$;  a large $B$ effectively
makes $\pi^\pm$ very massive and suppressed, letting the mass term, $f_\pi^2 m_\pi^2 (1-\cos f)$ insignificant in the energy function.


We expect that the lattice-QCD simulation is sufficiently capable of testing our theoretical prediction for the nucleon mass behavior in the left panel in Fig.~\ref{fig:Bdiff}.
From the observed magnetic moments, it is trivial that the masses should decrease for small $B$ (as long as our identification of the ground state as $|p\!\downarrow\rangle$ and the first excited state as $|n\!\uparrow\rangle$ is correct which implicitly assumes the full polarization).
It is not surprising to have rising $M$ for larger $B$ from the magnetic catalysis~\cite{Miransky:2015ava} that would generate heavier constituent quark masses.
We note that a similar trend was reported in the preceding Skyrmion calculation~\cite{He:2016oqk} under a simpler Ansatz.
It would be technically far more challenging to compute the pressure distribution but it should be in principle feasible by means of the gradient flow.  Then, whether the lattice-QCD results satisfy our proposed rum rule in the presence of the magnetic field would be directly examined.

Another interesting direction is, as exploited in Ref.~\cite{Fukushima:2020cmk}, the pressure distribution can be translated to the equation of state (EOS) i.e., a relation between the pressure and the energy density.  The EOS determined in this way was found to be fairly consistent with the empirically acceptable one in neutron stars~\cite{Fukushima:2020cmk}.  Our results in Fig.~\ref{fig:pressure} could be also converted in terms of the quark-core EOS, making a prediction for the magnetic field dependence.

This work was partially supported by JSPS KAKENHI Grant Nos.\ 21J20877 (S.C.), 202020974~PD (Z.Q.), 19K21874 (K.F.), 22H01216 (K.F.), and 22H05118 (K.F.).

\bibliographystyle{apsrev4-1}
\bibliography{skyrmion}

\begin{thebibliography}{31}%
\makeatletter
\providecommand \@ifxundefined [1]{%
 \@ifx{#1\undefined}
}%
\providecommand \@ifnum [1]{%
 \ifnum #1\expandafter \@firstoftwo
 \else \expandafter \@secondoftwo
 \fi
}%
\providecommand \@ifx [1]{%
 \ifx #1\expandafter \@firstoftwo
 \else \expandafter \@secondoftwo
 \fi
}%
\providecommand \natexlab [1]{#1}%
\providecommand \enquote  [1]{``#1''}%
\providecommand \bibnamefont  [1]{#1}%
\providecommand \bibfnamefont [1]{#1}%
\providecommand \citenamefont [1]{#1}%
\providecommand \href@noop [0]{\@secondoftwo}%
\providecommand \href [0]{\begingroup \@sanitize@url \@href}%
\providecommand \@href[1]{\@@startlink{#1}\@@href}%
\providecommand \@@href[1]{\endgroup#1\@@endlink}%
\providecommand \@sanitize@url [0]{\catcode `\\12\catcode `\$12\catcode
  `\&12\catcode `\#12\catcode `\^12\catcode `\_12\catcode `\%12\relax}%
\providecommand \@@startlink[1]{}%
\providecommand \@@endlink[0]{}%
\providecommand \url  [0]{\begingroup\@sanitize@url \@url }%
\providecommand \@url [1]{\endgroup\@href {#1}{\urlprefix }}%
\providecommand \urlprefix  [0]{URL }%
\providecommand \Eprint [0]{\href }%
\providecommand \doibase [0]{http://dx.doi.org/}%
\providecommand \selectlanguage [0]{\@gobble}%
\providecommand \bibinfo  [0]{\@secondoftwo}%
\providecommand \bibfield  [0]{\@secondoftwo}%
\providecommand \translation [1]{[#1]}%
\providecommand \BibitemOpen [0]{}%
\providecommand \bibitemStop [0]{}%
\providecommand \bibitemNoStop [0]{.\EOS\space}%
\providecommand \EOS [0]{\spacefactor3000\relax}%
\providecommand \BibitemShut  [1]{\csname bibitem#1\endcsname}%
\let\auto@bib@innerbib\@empty
\bibitem [{\citenamefont {Bogner}\ \emph {et~al.}(2010)\citenamefont {Bogner},
  \citenamefont {Furnstahl},\ and\ \citenamefont {Schwenk}}]{Bogner:2009bt}%
  \BibitemOpen
  \bibfield  {author} {\bibinfo {author} {\bibfnamefont {S.~K.}\ \bibnamefont
  {Bogner}}, \bibinfo {author} {\bibfnamefont {R.~J.}\ \bibnamefont
  {Furnstahl}}, \ and\ \bibinfo {author} {\bibfnamefont {A.}~\bibnamefont
  {Schwenk}},\ }\href {\doibase 10.1016/j.ppnp.2010.03.001} {\bibfield
  {journal} {\bibinfo  {journal} {Prog. Part. Nucl. Phys.}\ }\textbf {\bibinfo
  {volume} {65}},\ \bibinfo {pages} {94} (\bibinfo {year} {2010})},\ \Eprint
  {http://arxiv.org/abs/0912.3688} {arXiv:0912.3688 [nucl-th]} \BibitemShut
  {NoStop}%
\bibitem [{\citenamefont {'t~Hooft}(1974)}]{tHooft:1973alw}%
  \BibitemOpen
  \bibfield  {author} {\bibinfo {author} {\bibfnamefont {G.}~\bibnamefont
  {'t~Hooft}},\ }\href {\doibase 10.1016/0550-3213(74)90154-0} {\bibfield
  {journal} {\bibinfo  {journal} {Nucl. Phys. B}\ }\textbf {\bibinfo {volume}
  {72}},\ \bibinfo {pages} {461} (\bibinfo {year} {1974})}\BibitemShut
  {NoStop}%
\bibitem [{\citenamefont {Witten}(1979)}]{Witten:1979kh}%
  \BibitemOpen
  \bibfield  {author} {\bibinfo {author} {\bibfnamefont {E.}~\bibnamefont
  {Witten}},\ }\href {\doibase 10.1016/0550-3213(79)90232-3} {\bibfield
  {journal} {\bibinfo  {journal} {Nucl. Phys. B}\ }\textbf {\bibinfo {volume}
  {160}},\ \bibinfo {pages} {57} (\bibinfo {year} {1979})}\BibitemShut
  {NoStop}%
\bibitem [{\citenamefont {Witten}(1983{\natexlab{a}})}]{Witten:1983tw}%
  \BibitemOpen
  \bibfield  {author} {\bibinfo {author} {\bibfnamefont {E.}~\bibnamefont
  {Witten}},\ }\href {\doibase 10.1016/0550-3213(83)90063-9} {\bibfield
  {journal} {\bibinfo  {journal} {Nucl. Phys. B}\ }\textbf {\bibinfo {volume}
  {223}},\ \bibinfo {pages} {422} (\bibinfo {year}
  {1983}{\natexlab{a}})}\BibitemShut {NoStop}%
\bibitem [{\citenamefont {Witten}(1983{\natexlab{b}})}]{Witten:1983tx}%
  \BibitemOpen
  \bibfield  {author} {\bibinfo {author} {\bibfnamefont {E.}~\bibnamefont
  {Witten}},\ }\href {\doibase 10.1016/0550-3213(83)90064-0} {\bibfield
  {journal} {\bibinfo  {journal} {Nucl. Phys. B}\ }\textbf {\bibinfo {volume}
  {223}},\ \bibinfo {pages} {433} (\bibinfo {year}
  {1983}{\natexlab{b}})}\BibitemShut {NoStop}%
\bibitem [{\citenamefont {Skyrme}(1961)}]{Skyrme:1961vq}%
  \BibitemOpen
  \bibfield  {author} {\bibinfo {author} {\bibfnamefont {T.~H.~R.}\
  \bibnamefont {Skyrme}},\ }\href {\doibase 10.1098/rspa.1961.0018} {\bibfield
  {journal} {\bibinfo  {journal} {Proc. Roy. Soc. Lond. A}\ }\textbf {\bibinfo
  {volume} {260}},\ \bibinfo {pages} {127} (\bibinfo {year}
  {1961})}\BibitemShut {NoStop}%
\bibitem [{\citenamefont {Skyrme}(1962)}]{Skyrme:1962vh}%
  \BibitemOpen
  \bibfield  {author} {\bibinfo {author} {\bibfnamefont {T.~H.~R.}\
  \bibnamefont {Skyrme}},\ }\href {\doibase 10.1016/0029-5582(62)90775-7}
  {\bibfield  {journal} {\bibinfo  {journal} {Nucl. Phys.}\ }\textbf {\bibinfo
  {volume} {31}},\ \bibinfo {pages} {556} (\bibinfo {year} {1962})}\BibitemShut
  {NoStop}%
\bibitem [{\citenamefont {Miransky}\ and\ \citenamefont
  {Shovkovy}(2015)}]{Miransky:2015ava}%
  \BibitemOpen
  \bibfield  {author} {\bibinfo {author} {\bibfnamefont {V.~A.}\ \bibnamefont
  {Miransky}}\ and\ \bibinfo {author} {\bibfnamefont {I.~A.}\ \bibnamefont
  {Shovkovy}},\ }\href {\doibase 10.1016/j.physrep.2015.02.003} {\bibfield
  {journal} {\bibinfo  {journal} {Phys. Rept.}\ }\textbf {\bibinfo {volume}
  {576}},\ \bibinfo {pages} {1} (\bibinfo {year} {2015})},\ \Eprint
  {http://arxiv.org/abs/1503.00732} {arXiv:1503.00732 [hep-ph]} \BibitemShut
  {NoStop}%
\bibitem [{\citenamefont {Skokov}\ \emph {et~al.}(2009)\citenamefont {Skokov},
  \citenamefont {Illarionov},\ and\ \citenamefont {Toneev}}]{Skokov:2009qp}%
  \BibitemOpen
  \bibfield  {author} {\bibinfo {author} {\bibfnamefont {V.}~\bibnamefont
  {Skokov}}, \bibinfo {author} {\bibfnamefont {A.~Y.}\ \bibnamefont
  {Illarionov}}, \ and\ \bibinfo {author} {\bibfnamefont {V.}~\bibnamefont
  {Toneev}},\ }\href {\doibase 10.1142/S0217751X09047570} {\bibfield  {journal}
  {\bibinfo  {journal} {Int. J. Mod. Phys. A}\ }\textbf {\bibinfo {volume}
  {24}},\ \bibinfo {pages} {5925} (\bibinfo {year} {2009})},\ \Eprint
  {http://arxiv.org/abs/0907.1396} {arXiv:0907.1396 [nucl-th]} \BibitemShut
  {NoStop}%
\bibitem [{\citenamefont {Deng}\ and\ \citenamefont
  {Huang}(2012)}]{Deng:2012pc}%
  \BibitemOpen
  \bibfield  {author} {\bibinfo {author} {\bibfnamefont {W.-T.}\ \bibnamefont
  {Deng}}\ and\ \bibinfo {author} {\bibfnamefont {X.-G.}\ \bibnamefont
  {Huang}},\ }\href {\doibase 10.1103/PhysRevC.85.044907} {\bibfield  {journal}
  {\bibinfo  {journal} {Phys. Rev. C}\ }\textbf {\bibinfo {volume} {85}},\
  \bibinfo {pages} {044907} (\bibinfo {year} {2012})},\ \Eprint
  {http://arxiv.org/abs/1201.5108} {arXiv:1201.5108 [nucl-th]} \BibitemShut
  {NoStop}%
\bibitem [{\citenamefont {McLerran}\ and\ \citenamefont
  {Skokov}(2014)}]{McLerran:2013hla}%
  \BibitemOpen
  \bibfield  {author} {\bibinfo {author} {\bibfnamefont {L.}~\bibnamefont
  {McLerran}}\ and\ \bibinfo {author} {\bibfnamefont {V.}~\bibnamefont
  {Skokov}},\ }\href {\doibase 10.1016/j.nuclphysa.2014.05.008} {\bibfield
  {journal} {\bibinfo  {journal} {Nucl. Phys. A}\ }\textbf {\bibinfo {volume}
  {929}},\ \bibinfo {pages} {184} (\bibinfo {year} {2014})},\ \Eprint
  {http://arxiv.org/abs/1305.0774} {arXiv:1305.0774 [hep-ph]} \BibitemShut
  {NoStop}%
\bibitem [{\citenamefont {Kharzeev}\ \emph {et~al.}(2008)\citenamefont
  {Kharzeev}, \citenamefont {McLerran},\ and\ \citenamefont
  {Warringa}}]{Kharzeev:2007jp}%
  \BibitemOpen
  \bibfield  {author} {\bibinfo {author} {\bibfnamefont {D.~E.}\ \bibnamefont
  {Kharzeev}}, \bibinfo {author} {\bibfnamefont {L.~D.}\ \bibnamefont
  {McLerran}}, \ and\ \bibinfo {author} {\bibfnamefont {H.~J.}\ \bibnamefont
  {Warringa}},\ }\href {\doibase 10.1016/j.nuclphysa.2008.02.298} {\bibfield
  {journal} {\bibinfo  {journal} {Nucl. Phys. A}\ }\textbf {\bibinfo {volume}
  {803}},\ \bibinfo {pages} {227} (\bibinfo {year} {2008})},\ \Eprint
  {http://arxiv.org/abs/0711.0950} {arXiv:0711.0950 [hep-ph]} \BibitemShut
  {NoStop}%
\bibitem [{\citenamefont {Fukushima}\ \emph {et~al.}(2008)\citenamefont
  {Fukushima}, \citenamefont {Kharzeev},\ and\ \citenamefont
  {Warringa}}]{Fukushima:2008xe}%
  \BibitemOpen
  \bibfield  {author} {\bibinfo {author} {\bibfnamefont {K.}~\bibnamefont
  {Fukushima}}, \bibinfo {author} {\bibfnamefont {D.~E.}\ \bibnamefont
  {Kharzeev}}, \ and\ \bibinfo {author} {\bibfnamefont {H.~J.}\ \bibnamefont
  {Warringa}},\ }\href {\doibase 10.1103/PhysRevD.78.074033} {\bibfield
  {journal} {\bibinfo  {journal} {Phys. Rev. D}\ }\textbf {\bibinfo {volume}
  {78}},\ \bibinfo {pages} {074033} (\bibinfo {year} {2008})},\ \Eprint
  {http://arxiv.org/abs/0808.3382} {arXiv:0808.3382 [hep-ph]} \BibitemShut
  {NoStop}%
\bibitem [{\citenamefont {Kharzeev}\ \emph {et~al.}(2016)\citenamefont
  {Kharzeev}, \citenamefont {Liao}, \citenamefont {Voloshin},\ and\
  \citenamefont {Wang}}]{Kharzeev:2015znc}%
  \BibitemOpen
  \bibfield  {author} {\bibinfo {author} {\bibfnamefont {D.~E.}\ \bibnamefont
  {Kharzeev}}, \bibinfo {author} {\bibfnamefont {J.}~\bibnamefont {Liao}},
  \bibinfo {author} {\bibfnamefont {S.~A.}\ \bibnamefont {Voloshin}}, \ and\
  \bibinfo {author} {\bibfnamefont {G.}~\bibnamefont {Wang}},\ }\href {\doibase
  10.1016/j.ppnp.2016.01.001} {\bibfield  {journal} {\bibinfo  {journal} {Prog.
  Part. Nucl. Phys.}\ }\textbf {\bibinfo {volume} {88}},\ \bibinfo {pages} {1}
  (\bibinfo {year} {2016})},\ \Eprint {http://arxiv.org/abs/1511.04050}
  {arXiv:1511.04050 [hep-ph]} \BibitemShut {NoStop}%
\bibitem [{\citenamefont {Brauner}\ and\ \citenamefont
  {Yamamoto}(2017)}]{Brauner:2016pko}%
  \BibitemOpen
  \bibfield  {author} {\bibinfo {author} {\bibfnamefont {T.}~\bibnamefont
  {Brauner}}\ and\ \bibinfo {author} {\bibfnamefont {N.}~\bibnamefont
  {Yamamoto}},\ }\href {\doibase 10.1007/JHEP04(2017)132} {\bibfield  {journal}
  {\bibinfo  {journal} {JHEP}\ }\textbf {\bibinfo {volume} {04}},\ \bibinfo
  {pages} {132} (\bibinfo {year} {2017})},\ \Eprint
  {http://arxiv.org/abs/1609.05213} {arXiv:1609.05213 [hep-ph]} \BibitemShut
  {NoStop}%
\bibitem [{\citenamefont {Brauner}\ \emph {et~al.}(2019)\citenamefont
  {Brauner}, \citenamefont {Filios},\ and\ \citenamefont
  {Kole\v{s}ov\'a}}]{Brauner:2019rjg}%
  \BibitemOpen
  \bibfield  {author} {\bibinfo {author} {\bibfnamefont {T.}~\bibnamefont
  {Brauner}}, \bibinfo {author} {\bibfnamefont {G.}~\bibnamefont {Filios}}, \
  and\ \bibinfo {author} {\bibfnamefont {H.}~\bibnamefont {Kole\v{s}ov\'a}},\
  }\href {\doibase 10.1103/PhysRevLett.123.012001} {\bibfield  {journal}
  {\bibinfo  {journal} {Phys. Rev. Lett.}\ }\textbf {\bibinfo {volume} {123}},\
  \bibinfo {pages} {012001} (\bibinfo {year} {2019})},\ \Eprint
  {http://arxiv.org/abs/1902.07522} {arXiv:1902.07522 [hep-ph]} \BibitemShut
  {NoStop}%
\bibitem [{\citenamefont {D'Elia}\ \emph {et~al.}(2010)\citenamefont {D'Elia},
  \citenamefont {Mukherjee},\ and\ \citenamefont {Sanfilippo}}]{DElia:2010abb}%
  \BibitemOpen
  \bibfield  {author} {\bibinfo {author} {\bibfnamefont {M.}~\bibnamefont
  {D'Elia}}, \bibinfo {author} {\bibfnamefont {S.}~\bibnamefont {Mukherjee}}, \
  and\ \bibinfo {author} {\bibfnamefont {F.}~\bibnamefont {Sanfilippo}},\
  }\href {\doibase 10.1103/PhysRevD.82.051501} {\bibfield  {journal} {\bibinfo
  {journal} {Phys. Rev. D}\ }\textbf {\bibinfo {volume} {82}},\ \bibinfo
  {pages} {051501} (\bibinfo {year} {2010})},\ \Eprint
  {http://arxiv.org/abs/1005.5365} {arXiv:1005.5365 [hep-lat]} \BibitemShut
  {NoStop}%
\bibitem [{\citenamefont {Bali}\ \emph {et~al.}(2012)\citenamefont {Bali},
  \citenamefont {Bruckmann}, \citenamefont {Endrodi}, \citenamefont {Fodor},
  \citenamefont {Katz}, \citenamefont {Krieg}, \citenamefont {Schafer},\ and\
  \citenamefont {Szabo}}]{Bali:2011qj}%
  \BibitemOpen
  \bibfield  {author} {\bibinfo {author} {\bibfnamefont {G.~S.}\ \bibnamefont
  {Bali}}, \bibinfo {author} {\bibfnamefont {F.}~\bibnamefont {Bruckmann}},
  \bibinfo {author} {\bibfnamefont {G.}~\bibnamefont {Endrodi}}, \bibinfo
  {author} {\bibfnamefont {Z.}~\bibnamefont {Fodor}}, \bibinfo {author}
  {\bibfnamefont {S.~D.}\ \bibnamefont {Katz}}, \bibinfo {author}
  {\bibfnamefont {S.}~\bibnamefont {Krieg}}, \bibinfo {author} {\bibfnamefont
  {A.}~\bibnamefont {Schafer}}, \ and\ \bibinfo {author} {\bibfnamefont
  {K.~K.}\ \bibnamefont {Szabo}},\ }\href {\doibase 10.1007/JHEP02(2012)044}
  {\bibfield  {journal} {\bibinfo  {journal} {JHEP}\ }\textbf {\bibinfo
  {volume} {02}},\ \bibinfo {pages} {044} (\bibinfo {year} {2012})},\ \Eprint
  {http://arxiv.org/abs/1111.4956} {arXiv:1111.4956 [hep-lat]} \BibitemShut
  {NoStop}%
\bibitem [{\citenamefont {Hidaka}\ and\ \citenamefont
  {Yamamoto}(2013)}]{Hidaka:2012mz}%
  \BibitemOpen
  \bibfield  {author} {\bibinfo {author} {\bibfnamefont {Y.}~\bibnamefont
  {Hidaka}}\ and\ \bibinfo {author} {\bibfnamefont {A.}~\bibnamefont
  {Yamamoto}},\ }\href {\doibase 10.1103/PhysRevD.87.094502} {\bibfield
  {journal} {\bibinfo  {journal} {Phys. Rev. D}\ }\textbf {\bibinfo {volume}
  {87}},\ \bibinfo {pages} {094502} (\bibinfo {year} {2013})},\ \Eprint
  {http://arxiv.org/abs/1209.0007} {arXiv:1209.0007 [hep-ph]} \BibitemShut
  {NoStop}%
\bibitem [{\citenamefont {Ding}\ \emph {et~al.}(2021)\citenamefont {Ding},
  \citenamefont {Li}, \citenamefont {Tomiya}, \citenamefont {Wang},\ and\
  \citenamefont {Zhang}}]{Ding:2020hxw}%
  \BibitemOpen
  \bibfield  {author} {\bibinfo {author} {\bibfnamefont {H.~T.}\ \bibnamefont
  {Ding}}, \bibinfo {author} {\bibfnamefont {S.~T.}\ \bibnamefont {Li}},
  \bibinfo {author} {\bibfnamefont {A.}~\bibnamefont {Tomiya}}, \bibinfo
  {author} {\bibfnamefont {X.~D.}\ \bibnamefont {Wang}}, \ and\ \bibinfo
  {author} {\bibfnamefont {Y.}~\bibnamefont {Zhang}},\ }\href {\doibase
  10.1103/PhysRevD.104.014505} {\bibfield  {journal} {\bibinfo  {journal}
  {Phys. Rev. D}\ }\textbf {\bibinfo {volume} {104}},\ \bibinfo {pages}
  {014505} (\bibinfo {year} {2021})},\ \Eprint
  {http://arxiv.org/abs/2008.00493} {arXiv:2008.00493 [hep-lat]} \BibitemShut
  {NoStop}%
\bibitem [{\citenamefont {Chen}\ \emph {et~al.}(2022)\citenamefont {Chen},
  \citenamefont {Fukushima},\ and\ \citenamefont {Qiu}}]{Chen:2021vou}%
  \BibitemOpen
  \bibfield  {author} {\bibinfo {author} {\bibfnamefont {S.}~\bibnamefont
  {Chen}}, \bibinfo {author} {\bibfnamefont {K.}~\bibnamefont {Fukushima}}, \
  and\ \bibinfo {author} {\bibfnamefont {Z.}~\bibnamefont {Qiu}},\ }\href
  {\doibase 10.1103/PhysRevD.105.L011502} {\bibfield  {journal} {\bibinfo
  {journal} {Phys. Rev. D}\ }\textbf {\bibinfo {volume} {105}},\ \bibinfo
  {pages} {L011502} (\bibinfo {year} {2022})},\ \Eprint
  {http://arxiv.org/abs/2104.11482} {arXiv:2104.11482 [hep-ph]} \BibitemShut
  {NoStop}%
\bibitem [{\citenamefont {Polyakov}\ and\ \citenamefont
  {Schweitzer}(2018)}]{Polyakov:2018zvc}%
  \BibitemOpen
  \bibfield  {author} {\bibinfo {author} {\bibfnamefont {M.~V.}\ \bibnamefont
  {Polyakov}}\ and\ \bibinfo {author} {\bibfnamefont {P.}~\bibnamefont
  {Schweitzer}},\ }\href {\doibase 10.1142/S0217751X18300259} {\bibfield
  {journal} {\bibinfo  {journal} {Int. J. Mod. Phys. A}\ }\textbf {\bibinfo
  {volume} {33}},\ \bibinfo {pages} {1830025} (\bibinfo {year} {2018})},\
  \Eprint {http://arxiv.org/abs/1805.06596} {arXiv:1805.06596 [hep-ph]}
  \BibitemShut {NoStop}%
\bibitem [{\citenamefont {Polyakov}\ and\ \citenamefont
  {Weiss}(1999)}]{Polyakov:1999gs}%
  \BibitemOpen
  \bibfield  {author} {\bibinfo {author} {\bibfnamefont {M.~V.}\ \bibnamefont
  {Polyakov}}\ and\ \bibinfo {author} {\bibfnamefont {C.}~\bibnamefont
  {Weiss}},\ }\href {\doibase 10.1103/PhysRevD.60.114017} {\bibfield  {journal}
  {\bibinfo  {journal} {Phys. Rev. D}\ }\textbf {\bibinfo {volume} {60}},\
  \bibinfo {pages} {114017} (\bibinfo {year} {1999})},\ \Eprint
  {http://arxiv.org/abs/hep-ph/9902451} {arXiv:hep-ph/9902451} \BibitemShut
  {NoStop}%
\bibitem [{\citenamefont {Gibbons}\ \emph {et~al.}(2011)\citenamefont
  {Gibbons}, \citenamefont {Warnick},\ and\ \citenamefont
  {Wong}}]{Gibbons:2010cr}%
  \BibitemOpen
  \bibfield  {author} {\bibinfo {author} {\bibfnamefont {G.~W.}\ \bibnamefont
  {Gibbons}}, \bibinfo {author} {\bibfnamefont {C.~M.}\ \bibnamefont
  {Warnick}}, \ and\ \bibinfo {author} {\bibfnamefont {W.~W.}\ \bibnamefont
  {Wong}},\ }\href {\doibase 10.1063/1.3523469} {\bibfield  {journal} {\bibinfo
   {journal} {J. Math. Phys.}\ }\textbf {\bibinfo {volume} {52}},\ \bibinfo
  {pages} {012905} (\bibinfo {year} {2011})},\ \Eprint
  {http://arxiv.org/abs/1005.2488} {arXiv:1005.2488 [hep-th]} \BibitemShut
  {NoStop}%
\bibitem [{\citenamefont {Burkert}\ \emph {et~al.}(2018)\citenamefont
  {Burkert}, \citenamefont {Elouadrhiri},\ and\ \citenamefont
  {Girod}}]{Burkert:2018bqq}%
  \BibitemOpen
  \bibfield  {author} {\bibinfo {author} {\bibfnamefont {V.}~\bibnamefont
  {Burkert}}, \bibinfo {author} {\bibfnamefont {L.}~\bibnamefont
  {Elouadrhiri}}, \ and\ \bibinfo {author} {\bibfnamefont {F.}~\bibnamefont
  {Girod}},\ }\href {\doibase 10.1038/s41586-018-0060-z} {\bibfield  {journal}
  {\bibinfo  {journal} {Nature}\ }\textbf {\bibinfo {volume} {557}},\ \bibinfo
  {pages} {396} (\bibinfo {year} {2018})}\BibitemShut {NoStop}%
\bibitem [{\citenamefont {Canfora}\ and\ \citenamefont
  {Maeda}(2013)}]{Canfora:2013osa}%
  \BibitemOpen
  \bibfield  {author} {\bibinfo {author} {\bibfnamefont {F.}~\bibnamefont
  {Canfora}}\ and\ \bibinfo {author} {\bibfnamefont {H.}~\bibnamefont
  {Maeda}},\ }\href {\doibase 10.1103/PhysRevD.87.084049} {\bibfield  {journal}
  {\bibinfo  {journal} {Phys. Rev. D}\ }\textbf {\bibinfo {volume} {87}},\
  \bibinfo {pages} {084049} (\bibinfo {year} {2013})},\ \Eprint
  {http://arxiv.org/abs/1302.3232} {arXiv:1302.3232 [gr-qc]} \BibitemShut
  {NoStop}%
\bibitem [{\citenamefont {Ayon-Beato}\ \emph {et~al.}(2016)\citenamefont
  {Ayon-Beato}, \citenamefont {Canfora},\ and\ \citenamefont
  {Zanelli}}]{Ayon-Beato:2015eca}%
  \BibitemOpen
  \bibfield  {author} {\bibinfo {author} {\bibfnamefont {E.}~\bibnamefont
  {Ayon-Beato}}, \bibinfo {author} {\bibfnamefont {F.}~\bibnamefont {Canfora}},
  \ and\ \bibinfo {author} {\bibfnamefont {J.}~\bibnamefont {Zanelli}},\ }\href
  {\doibase 10.1016/j.physletb.2015.11.065} {\bibfield  {journal} {\bibinfo
  {journal} {Phys. Lett. B}\ }\textbf {\bibinfo {volume} {752}},\ \bibinfo
  {pages} {201} (\bibinfo {year} {2016})},\ \Eprint
  {http://arxiv.org/abs/1509.02659} {arXiv:1509.02659 [gr-qc]} \BibitemShut
  {NoStop}%
\bibitem [{\citenamefont {Adkins}\ and\ \citenamefont
  {Nappi}(1984)}]{Adkins:1983hy}%
  \BibitemOpen
  \bibfield  {author} {\bibinfo {author} {\bibfnamefont {G.~S.}\ \bibnamefont
  {Adkins}}\ and\ \bibinfo {author} {\bibfnamefont {C.~R.}\ \bibnamefont
  {Nappi}},\ }\href {\doibase 10.1016/0550-3213(84)90172-X} {\bibfield
  {journal} {\bibinfo  {journal} {Nucl. Phys. B}\ }\textbf {\bibinfo {volume}
  {233}},\ \bibinfo {pages} {109} (\bibinfo {year} {1984})}\BibitemShut
  {NoStop}%
\bibitem [{\citenamefont {Ohtani}\ and\ \citenamefont
  {Ohta}(2004)}]{Ohtani:2004aw}%
  \BibitemOpen
  \bibfield  {author} {\bibinfo {author} {\bibfnamefont {M.}~\bibnamefont
  {Ohtani}}\ and\ \bibinfo {author} {\bibfnamefont {K.}~\bibnamefont {Ohta}},\
  }\href {\doibase 10.1103/PhysRevD.70.096014} {\bibfield  {journal} {\bibinfo
  {journal} {Phys. Rev. D}\ }\textbf {\bibinfo {volume} {70}},\ \bibinfo
  {pages} {096014} (\bibinfo {year} {2004})},\ \Eprint
  {http://arxiv.org/abs/hep-ph/0406173} {arXiv:hep-ph/0406173} \BibitemShut
  {NoStop}%
\bibitem [{\citenamefont {He}(2017)}]{He:2016oqk}%
  \BibitemOpen
  \bibfield  {author} {\bibinfo {author} {\bibfnamefont {B.-R.}\ \bibnamefont
  {He}},\ }\href {\doibase 10.1016/j.physletb.2016.12.019} {\bibfield
  {journal} {\bibinfo  {journal} {Phys. Lett. B}\ }\textbf {\bibinfo {volume}
  {765}},\ \bibinfo {pages} {109} (\bibinfo {year} {2017})},\ \Eprint
  {http://arxiv.org/abs/1609.09055} {arXiv:1609.09055 [hep-ph]} \BibitemShut
  {NoStop}%
\bibitem [{\citenamefont {Fukushima}\ \emph {et~al.}(2020)\citenamefont
  {Fukushima}, \citenamefont {Kojo},\ and\ \citenamefont
  {Weise}}]{Fukushima:2020cmk}%
  \BibitemOpen
  \bibfield  {author} {\bibinfo {author} {\bibfnamefont {K.}~\bibnamefont
  {Fukushima}}, \bibinfo {author} {\bibfnamefont {T.}~\bibnamefont {Kojo}}, \
  and\ \bibinfo {author} {\bibfnamefont {W.}~\bibnamefont {Weise}},\ }\href
  {\doibase 10.1103/PhysRevD.102.096017} {\bibfield  {journal} {\bibinfo
  {journal} {Phys. Rev. D}\ }\textbf {\bibinfo {volume} {102}},\ \bibinfo
  {pages} {096017} (\bibinfo {year} {2020})},\ \Eprint
  {http://arxiv.org/abs/2008.08436} {arXiv:2008.08436 [hep-ph]} \BibitemShut
  {NoStop}%
\end{thebibliography}%

\end{document}